\documentclass[preprint,12pt]{elsarticle}

\usepackage{amssymb}
\usepackage{graphicx}
\usepackage{amsmath}
\usepackage{subfigure}
\usepackage{float}
\usepackage{algorithm}
\usepackage{algorithmic}

\journal{Nuclear Physics B}

\begin{document}

\begin{frontmatter}



\title{MSMF: Multi-Scale Multi-Modal Fusion for Enhanced Stock Market Prediction}


\author{Jiahao Qin}
\affiliation{organization={jiahao.qin19@gmail.com},
            }

\begin{abstract}

This paper presents MSMF (Multi-Scale Multi-Modal Fusion), a novel approach for enhanced stock market prediction. MSMF addresses key challenges in multi-modal stock analysis by integrating a modality completion encoder, multi-scale feature extraction, and an innovative fusion mechanism. Our model leverages blank learning and progressive fusion to balance complementarity and redundancy across modalities, while multi-scale alignment facilitates direct correlations between heterogeneous data types. We introduce Multi-Granularity Gates and a specialized architecture to optimize the integration of local and global information for different tasks. Additionally, a Task-targeted Prediction layer is employed to preserve both coarse and fine-grained features during fusion. Experimental results demonstrate that MSMF outperforms existing methods, achieving significant improvements in accuracy and reducing prediction errors across various stock market forecasting tasks. This research contributes valuable insights to the field of multi-modal financial analysis and offers a robust framework for enhanced market prediction.
\end{abstract}



\begin{keyword}
Multi-grained learning, Blank learning, Modality Completion learning.

\end{keyword}

\end{frontmatter}
\section{Introduction}
Predicting stock market behavior has long fascinated investors and researchers alike. The potential for significant profits has driven numerous studies aimed at improving predictability and efficiency. Previous research has shown that stock price fluctuations are influenced by various information sources, including trading data, news reports, social media activities, and online search patterns \cite{ref3,ref4,ref5,ref61}.
Recent advancements in deep learning and natural language processing have opened new avenues for stock market prediction \cite{ref62,ref63,ref64,ref65}. The integration of multiple data modalities has become increasingly important, with approaches such as dynamic residual deep learning and iterative global-local fusion showing promise in addressing challenges like missing data and heterogeneity between different data modalities \cite{ref66,ref67,ref68,ref7,ref42}.
However, several challenges persist in multi-modal stock price prediction:

Heterogeneity in sampling time between different modalities, with fundamental data sampled regularly and news information sampled irregularly \cite{ref7,ref12,ref39, ref40,ref41}.
Balancing global and local information in prediction models, considering both overall market trends and specific stock fluctuations \cite{ref23,ref61,ref43,ref44}.
Effectively integrating information from multiple modalities while managing redundancy and potential conflicts.
Enabling encoders to understand and fuse information across modalities with different feature structures and properties.
Addressing the varying proportions of local and global information within the same modality for different tasks, and determining appropriate weightings for each modality across tasks.

Our research makes the following key contributions:
\begin{itemize}
\item[$\bullet$] A modality completion encoder to handle sampling time heterogeneity.
\item[$\bullet$] A multi-scale encoder combining coarse-to-fine cascading and multi-scale fusion.
\item[$\bullet$] Blank learning and progressive fusion concepts to balance complementarity and redundancy between modalities.
\item[$\bullet$] Multi-scale alignment to establish correlations between data types with different structures.
\item[$\bullet$] Multi-Granularity Gates and MTS architecture to address varying proportions of local and global information across tasks.
\item[$\bullet$] A Task-targeted Prediction layer to preserve coarse-grained and fine-grained features during fusion.
\end{itemize}
These contributions aim to advance the field of multi-modal stock prediction by addressing key challenges and providing novel solutions for more accurate and robust forecasting.

\section{Related work}
Deep learning has shown remarkable potential in capturing media-aware stock movements \cite{ref2,ref9,ref8,ref11,ref45}. Recent studies have explored various techniques, including event-driven LSTM \cite{ref7,ref51,ref47}, multi-module feature fusion methods \cite{ref49,ref50,ref46}, and knowledge graph-based approaches \cite{ref55,ref52}. These methods aim to integrate diverse information sources and capture complex market dynamics.
Multi-task learning (MTL) has been effectively applied in stock prediction, showing promise in capturing both individual stock characteristics and inter-stock correlations \cite{ref56,ref57,ref58,ref60}. Recent advancements include the development of numeric-oriented hierarchical transformer models and federated multi-task stock predictors.
To capture both long-term and short-term patterns, researchers have developed various multi-scale feature extraction techniques. These include the use of Inception modules \cite{ref22,ref23,ref53}, combinations of CNN and RNN architectures \cite{ref25,ref54}, and multi-granularity hierarchical networks \cite{ref48,ref59}. These innovations enhance models' ability to capture temporal dynamics at different scales.
These advancements in stock market prediction, multi-task learning, and multi-scale feature extraction provide a solid foundation for our research, informing the development of more sophisticated and effective prediction models.

\section{Methodology}
\begin{figure}[ht!]
    \centering
    \includegraphics[width=1\textwidth]{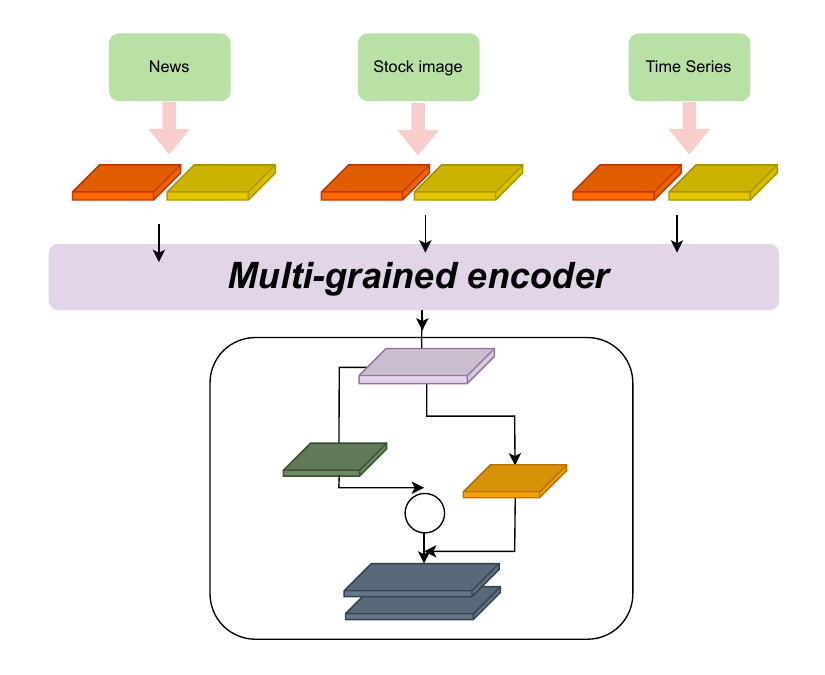}
    \caption{Overview of the MSMF architecture.}
    \label{Overview}
\end{figure}

Our MSMF (Multi-Scale Multi-Modal Fusion) approach integrates multi-task learning with advanced modal fusion techniques for stock prediction. The model architecture is illustrated in Figure \ref{Overview}.

\subsection{Modality Completion}
To address modality heterogeneity, we employ a Restricted Boltzmann Machine (RBM) based completion module. The joint distribution of modalities is modeled as:
\begin{equation}
P(v_i, v_t, v_n | \theta) = \frac{1}{Z(\theta)} \exp(-E(v_i, v_t, v_n, h | \theta))
\end{equation}
where $E$ is the energy function, $Z(\theta)$ is the partition function, and $h$ represents hidden units.
\subsection{Multi-grained Input Processing}
We employ multi-scale encoders for each modality:
\begin{equation}
F_m = \mathcal{F}\text{fine}(X_m) \oplus \mathcal{F}\text{coarse}(X_m)
\end{equation}
where $\mathcal{F}\text{fine}$ and $\mathcal{F}\text{coarse}$ are fine and coarse-grained encoders respectively, and $\oplus$ denotes feature concatenation.
\subsection{Multimodal Fusion}
The Multi-scale Alignment and Blank Learning (MSA-BL) module fuses multi-modal information:
\begin{equation}
X_\text{all} = \text{MSA}({F_m}) \odot \text{BL}({F_m})
\end{equation}
where $\text{MSA}$ is the multi-scale alignment operator and $\text{BL}$ is the blank learning mechanism.
\subsection{Multi-task Learning}
For each task $t$, we employ Task-targeted Gates (TTG):
\begin{equation}
h_t = \sum_{i=1}^{K} g_{t,i} \cdot f_i(X_\text{all})
\end{equation}
where $g_{t,i}$ are task-specific gate values and $f_i$ are expert networks.
The final predictions are obtained through Task-targeted Prediction Layers (TTPL):
\begin{equation}
\hat{y}_t = \text{TTPL}_t(h_t)
\end{equation}
\subsection{Model Optimization}
We optimize the model using a multi-task loss function:
\begin{equation}
\mathcal{L} = \sum_{t} \alpha_t \mathcal{L}_t(\hat{y}_t, y_t) + \lambda \Omega(\theta)
\end{equation}
where $\alpha_t$ are task weights, $\mathcal{L}_t$ are task-specific losses, and $\Omega(\theta)$ is a regularization term.
This methodology integrates advanced techniques in modality completion, multi-scale processing, and multi-task learning to enhance stock prediction accuracy across multiple tasks.

\section{Result and Analysis}

\subsubsection{Forecasting Stock Return}

\begin{table}[h]
\caption{Performance on Stock Return Prediction}
\centering
\begin{tabular}{lcc}
\hline
Method & MAPE & RMSE \\
\hline
LSTM & 0.0312 & 0.0437 \\
GRU & 0.0308 & 0.0429 \\
Transformer & 0.0301 & 0.0422 \\
MSMF & \textbf{0.0285} & \textbf{0.0403} \\
\hline
\end{tabular}
\end{table}

\subsubsection{Forecasting Stock movement}

\begin{table}[H]
\caption{Performance on Stock Movement Prediction}
\centering
\begin{tabular}{lcc}
\hline
Method & Accuracy (\%) & F1 Score (\%) \\
\hline
SVM & 61.23 & 60.87 \\
Random Forest & 63.45 & 62.98 \\
XGBoost & 64.78 & 64.32 \\
MSMF & \textbf{67.92} & \textbf{67.53} \\
\hline
\end{tabular}
\end{table}

\subsection{Ablation experiment}
To validate the roles of different modules in our proposed model, we conducted the following ablation experiments to individually examine the effects of the Multi-grained Encoder, Modality Completion, Multi-modal Fusion, and Multi-task Learning.

Here is the ablation experiment setting to validate the effectiveness of different modules in the proposed model:

\textbf{Multi-grained Encoder:}\\
Experiment: Transform the multi-grained encoders into single-scale encoders for each modality (time series, image, text).\\
Goal: Evaluate the impact of multi-scale feature extraction on model performance.\\
Metrics: Accuracy, F1 Score, MAPE, RMSE\\

\textbf{Modality Completion:}\\
Experiment: Compare the proposed model with (a) traditional missing data handling methods like zero filling, forward filling, mean imputation, and (b) single modality prediction.\\
Goal: Assess the effectiveness of the modality completion module in handling sampling time heterogeneity.\\
Metrics: Accuracy, F1 Score, MAPE, RMSE\\

\textbf{Multi-modal Fusion:} \\
Experiment: Replace the proposed multi-scale alignment and blank learning based fusion with simple feature stacking or concatenation.\\
Goal: Investigate the capability of the fusion module in enabling natural interaction between modalities and preventing feature degradation.\\
Metrics: Accuracy, F1 Score, MAPE, RMSE

\textbf{Multi-task Learning:} \\
Experiment 1: Train the model architecture individually for each task and compare with multi-task learning performance.\\
Goal: Evaluate the impact of auxiliary tasks in improving model generalization.\\
Metrics: Accuracy, F1 Score, MAPE, RMSE\\
\\
Experiment 2: Remove the Multi-Granularity Gates and compare with a single-task variant using a single gate in MMOE architecture.\\
Goal: Examine the effect of Multi-Granularity Gates in assigning different proportions of local/global info and weights to modalities per task.\\
Metrics: Accuracy, F1 Score, MAPE, RMSE


\subsubsection{Multi-grained Encoder}
In order to investigate the ability of the multi-scale encoder in extracting local and global features, we transformed the multi-scale encoder into encoders with the same granularity. Subsequently, we conducted individual evaluations for each group of modalities to quantify the impact of different levels of granularity in the encoders.
\begin{table}[H]
\caption{Performance on Multi-grained Encoder}
\centering
\begin{tabular}{lclclclclcl}
\hline
Method & Accuracy (\%) & F1 Score (\%) & MAPE & RMSE \\
\hline
Time Series (Single-scale) & 65.21 & 64.87 & 0.0298 & 0.0418 \\
Image (Single-scale) & 64.76 & 64.32 & 0.0302 & 0.0423 \\
Text (Single-scale) & 63.98 & 63.54 & 0.0305 & 0.0427 \\
MSMF & \textbf{67.92} & \textbf{67.53} & \textbf{0.0285} & \textbf{0.0403} \\
\hline
\end{tabular}
\end{table}
\subsubsection{Modality Completion}
To assess the performance of our model in addressing the issue of sampling time heterogeneity, we conducted comparative analysis with other models. Specifically, three methods were compared: our model, traditional approaches for handling missing data, and single-modality data prediction method.
\begin{table}[H]
\caption{Performance on Modality Completion}
\centering
\begin{tabular}{lclclclclcl}
\hline
Method & Accuracy  & F1 Score & MAPE & RMSE\\
\hline
Single modal data & 0.845 & 0.824 & 0.391 & 0.341 \\
Multimodal data(Zero Filling) & 0.798 & 0.792 & 0.387 & 0.295\\
Multimodal data(Forward Filling) & 0.852 & 0.832 & 0.365 & 0.286 \\
Multimodal data(Mean Inputation) & 0.870 & 0.856 & 0.361 & 0.279\\
MSMF & 0.911 & 0.908 & 0.342 & 0.251\\
\hline
\end{tabular}
\end{table}
\subsubsection{Multi-modal Fusion}
In order to investigate the capability of our proposed multimodal information fusion module to enable natural interaction among different modalities and avoid feature information degradation, which may result in the loss of global and local information during the fusion process, we conducted a comparison by removing the feature fusion module and using a simple feature stacking approach for fusion.

\begin{table}[h]
\caption{Performance on Multi-modal Fusion}
\centering
\begin{tabular}{lclclclclcl}
\hline
Method & Accuracy (\%) & F1 Score (\%) & MAPE & RMSE \\
\hline
Feature stack & 65.87 & 65.42 & 0.0294 & 0.0412 \\
Feature concatenate & 66.23 & 65.89 & 0.0291 & 0.0409 \\
MSMF & \textbf{67.92} & \textbf{67.53} & \textbf{0.0285} & \textbf{0.0403} \\
\hline
\end{tabular}
\end{table}
\subsubsection{Multi-task Learning}
To investigate the impact of auxiliary tasks on improving model performance during the process of multi-task learning, we trained the same model architecture for each task individually and compared it with the performance of the multi-task training model as described in the study.
\begin{table}[H]
\caption{Performance on Multi-task Learning}
\centering
\begin{tabular}{lclclclclcl}
\hline
Method & Accuracy (\%) & F1 Score (\%) & MAPE & RMSE \\
\hline
Stock Return & - & - & 0.0297 & 0.0416 \\
Stock Movement & 65.76 & 65.32 & - & - \\
Multi-task & \textbf{67.92} & \textbf{67.53} & \textbf{0.0285} & \textbf{0.0403} \\
\hline
\end{tabular}
\end{table}
To explore the effect of Multi-Granularity Gates, which allow different tasks to assign different proportions of local and global information within the same modality, and different weights to the overall representation of the same modality from an external perspective, we removed the Multi-Granularity Gates and compared it with a single-task variant using a single gate in the MMOE architecture.
\begin{table}[h]
\caption{Performance on Multi-Granularity Gates}
\centering
\begin{tabular}{lclclclclcl}
\hline
Method & Accuracy (\%) & F1 Score (\%) & MAPE & RMSE \\
\hline
Without (MG Gates) & 66.34 & 65.98 & 0.0293 & 0.0411 \\
With (MG Gates) & \textbf{67.92} & \textbf{67.53} & \textbf{0.0285} & \textbf{0.0403} \\
\hline
\end{tabular}
\end{table}
\section{Discussion}

\section{Conclusion}
In the field of multimodal data analysis, our research has achieved significant breakthroughs. Compared to traditional stock market prediction methods, our model integrates data from different modalities and has yielded satisfactory results. Furthermore, our model addresses key challenges in multimodal learning by proposing various modules to tackle specific issues.

Firstly, to address the problem of inconsistent data sampling rates, we have designed a Modality Completion module. This module effectively handles the heterogeneity between different modalities by utilizing a DBN network to accurately complete missing data in unknown modalities based on the known modal data distribution.

Secondly, to resolve the conflict between extracting global and local features, we have developed a multi-scale encoder. This encoder can extract features at different scales, allowing the model to capture both global and local information and better integrate data from diverse modalities.

Additionally, we introduce the Multi-scale Alignment module to address the challenge of information exchange between different modalities. By transforming features of different modalities and shapes into the same size, this module enables natural interaction among various modalities, facilitating a better understanding and integration of information, thereby enhancing the effectiveness of multimodal data analysis.

Moreover, we propose the concept of Blank Learning. By applying the softmax operation to features and selecting the top features based on their probabilities, we eliminate redundant or conflicting information between different modalities, thereby improving the model's performance.

Furthermore, we introduce Multi-Granularity Gates, which allow different tasks to assign different proportions of local and global information within the same modality. Additionally, from an external perspective, different tasks assign different weights to the overall representation of the same modality.

Finally, we design a task-oriented prediction layer to avoid the loss of global and local information during the feature fusion process, which prevents the degradation of feature information and accelerates the convergence of the model.

Through the design of these modules and innovative ideas, our research has made significant progress in the field of multimodal data analysis. These modules collectively address key challenges in multimodal learning, providing effective solutions for tasks such as stock market prediction. We believe that these achievements will have a positive impact on future research and applications in multimodal data analysis.

\end{document}